\documentclass[runningheads]{llncs}
\usepackage{appendix}
\usepackage{graphicx}
\usepackage{comment}
\usepackage{booktabs}
\usepackage{subcaption}
\usepackage{xcolor}
\usepackage{mwe}

\makeatletter
\newcommand{\printfnsymbol}[1]{%
  \textsuperscript{\@fnsymbol{#1}}%
}
\makeatother

\usepackage{amsmath,amsfonts,bm}

\newcommand{\E}{\mathbb{E}}

\newcommand{\x}{\mathbf{x}}

\newcommand{\calL}{\mathcal{L}}

\newcommand{\RN}[1]{%
  \textup{\uppercase\expandafter{\romannumeral#1}}%
}

\begin{document}

\title{Mitigating Calibration Bias Without Fixed Attribute Grouping for Improved Fairness in Medical Imaging Analysis }

\titlerunning{Calibration Bias Mitigation}
%
\author{Changjian Shui \inst{1,2,\thanks{Equal contribution}} \and Justin Szeto \inst{1,2,\printfnsymbol{1}}  \and Raghav Mehta \inst{1,2} \and Douglas L. Arnold \inst{3,4} \and Tal Arbel\inst{1,2}}
\authorrunning{Shui et al.}
%
\institute{Center for Intelligent Machines, McGill University, Canada\\
\and MILA, Quebec AI Institute, Montreal, Canada\\
\and Department of Neurology and Neurosurgery, McGill University, Canada \\
\and NeuroRx Research, Montreal, Canada \\ \email{\{maxshui,jszeto,raghav,arbel\}@cim.mcgill.ca}, \email{douglas.arnold@mcgill.ca}}
\maketitle  
\begin{abstract}


Trustworthy deployment of deep learning medical imaging models into real-world clinical practice requires that they be calibrated. However, models that are well calibrated overall can still be poorly calibrated for a sub-population, potentially resulting in a clinician unwittingly making poor decisions for this group based on the recommendations of the model. Although methods have been shown to successfully mitigate biases across subgroups in terms of model accuracy, this work focuses on the open problem of mitigating calibration biases in the context of medical image analysis. Our method does not require subgroup attributes during training, permitting the flexibility to mitigate biases for different choices of sensitive attributes without re-training. To this end, we propose a novel two-stage method: Cluster-Focal to first identify poorly calibrated samples, cluster them into groups, and then introduce group-wise focal loss to improve calibration bias. We evaluate our method on skin lesion classification with the public HAM10000 dataset, and on predicting future lesional activity for multiple sclerosis (MS) patients. In addition to considering traditional sensitive attributes (e.g. age, sex) with demographic subgroups, we also consider biases among groups with different image-derived attributes, such as lesion load, which are required in medical image analysis.  Our results demonstrate that our method effectively controls calibration error in the worst-performing subgroups while preserving prediction performance, and outperforming recent baselines.

\end{abstract}

\section{Introduction}

\begin{figure}[t]
\begin{subfigure}{.5\textwidth}
  \centering
  \includegraphics[width=0.95\linewidth]{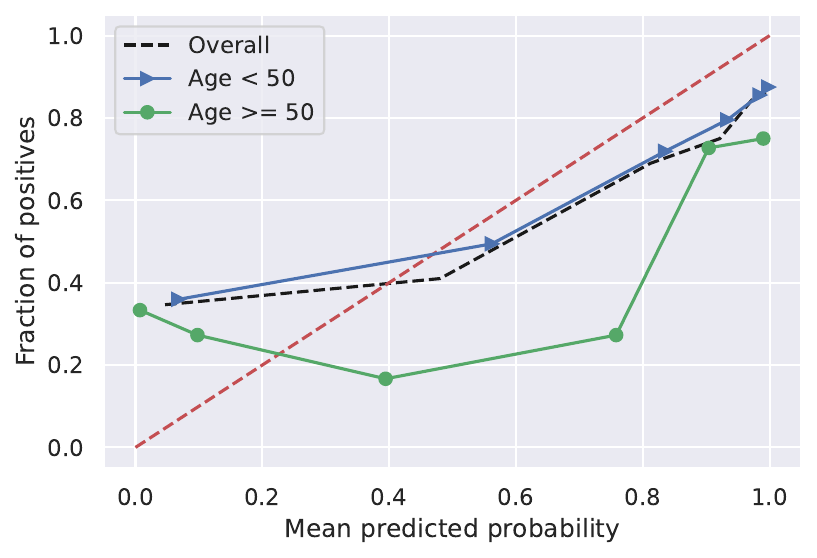}
  \caption{Reliability diagram: \textsf{ERM}}
\end{subfigure}%
\begin{subfigure}{.5\textwidth}
  \centering
  \includegraphics[width=.90\linewidth]{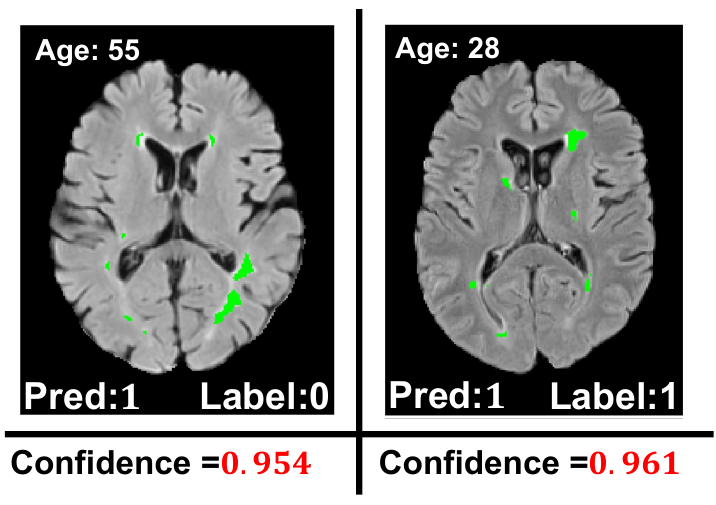}
  \caption{Example Calibration Results: MS}
\end{subfigure}
\caption{Illustration of calibration bias for a model that predicts future new lesional activity for multiple sclerosis (MS) patients. (a) Reliability diagram: \textsf{ERM} (training without considering any fairness) exhibits good calibration overall and also for younger patients, whereas it produces significantly overconfident and incorrect predictions for older patients. (b) Two MS patients  depicting highly confident predictions, with incorrect results on the older patient and correct results on the younger patient. Poorer calibration for older patients results in older patients being more likely to be incorrect with high confidence.}
\label{fig:motivation}
\end{figure}

Deep learning models have shown high prediction performance on many medical imaging tasks (e.g.,\cite{codella2018skin,6975210,10.1007/978-3-030-11723-8_6,pmlr-v102-tousignant19a}). 
However, deep learning models can indeed make errors, leading to distrust and hesitation by clinicians to integrate them into their workflows. In particular, models that show a tendency for overconfident incorrect predictions present real risk to patient care if deployed in real clinical practice. One way to improve the trustworthiness of a model is to ensure that it is well-calibrated, in that the predicted probabilities of the outcomes align with the probability of making a correct prediction~\cite{guo2017calibration}. While several methods have been shown to successfully improve calibration on the \emph{overall} population~\cite{guo2017calibration,mukhoti2020calibrating}, they cannot guarantee a small calibration error on \emph{sub-populations}. This can lead to a lack of fairness and equity in the resulting diagnostic decisions for a subset of the population.  Figure~\ref{fig:motivation}(a) illustrates how a deep learning model can achieve good calibration for the overall population and for younger patients, but produces significantly overconfident and incorrect predictions for older patients.  




\noindent Although various methods have been shown to successfully mitigate biases by improving prediction performance (e.g. accuracy) in the worst-performing subgroup~\cite{zou2018ai,larrazabal2020gender,ricci2022addressing,burlina2021addressing,zong2023}, improved prediction performance does not necessarily imply better calibration. As such, this paper focuses on the open problem of mitigating calibration bias in medical image analysis. Moreover, our method does not require subgroup attributes during the training, which permits the flexibility to mitigate biases for different choices of sensitive attributes without re-training. 

\noindent This paper proposes a novel two-stage method:~\textsf{Cluster-Focal}. In the first stage, a model $f_{id}$ is trained to identify poorly calibrated samples. The samples are then clustered according to their calibration gap. In the next stage, a prediction model $f_{\text{pred}}$ is trained via group-wise focal loss. Extensive experiments are performed on (a) skin lesion classification, based on the public HAM10000 dataset~\cite{codella2018skin}, and (b) on predicting future new lesional activity for multiple sclerosis (MS) patients on a proprietary, federated dataset of MRI acquired during  different clinical trials ~\cite{ms_bravo_trial,ms_advance_trial,ms_define_trial}. At test time, calibration bias mitigation is examined on subgroups based on sensitive demographic attributes (e.g. age, sex). In addition, we consider subgroups with different image-derived attributes, such as lesion load.
We further compare \textsf{Cluster-Focal} with recent debiasing methods that do not need subgroup annotations, such as EIIL (Environment Inference for Invariant Learning)~\cite{creager2021environment}, ARL (Adversarially Reweighted Learning)~\cite{NEURIPS2020_07fc15c9}, and JTT (Just Train Twice)~\cite{liu2021just}. Results demonstrate that \textsf{Cluster-Focal} can effectively reduce calibration error in the worst-performing subgroup, while preserving good prediction performance, when split into different subgroups based on a variety of attributes.




\section{Methodology}
 We propose a two-stage training strategy, \textsf{Cluster-Focal}. The first stage consists of \emph{identifying different levels of poorly calibrated samples}. In the second stage, we introduce a group-wise focal loss to mitigate the calibration bias. At test time, our model can mitigate biases for a variety of relevant subgroups of interest.

\begin{figure}[t]
  \centering
  \includegraphics[width=0.8\textwidth]{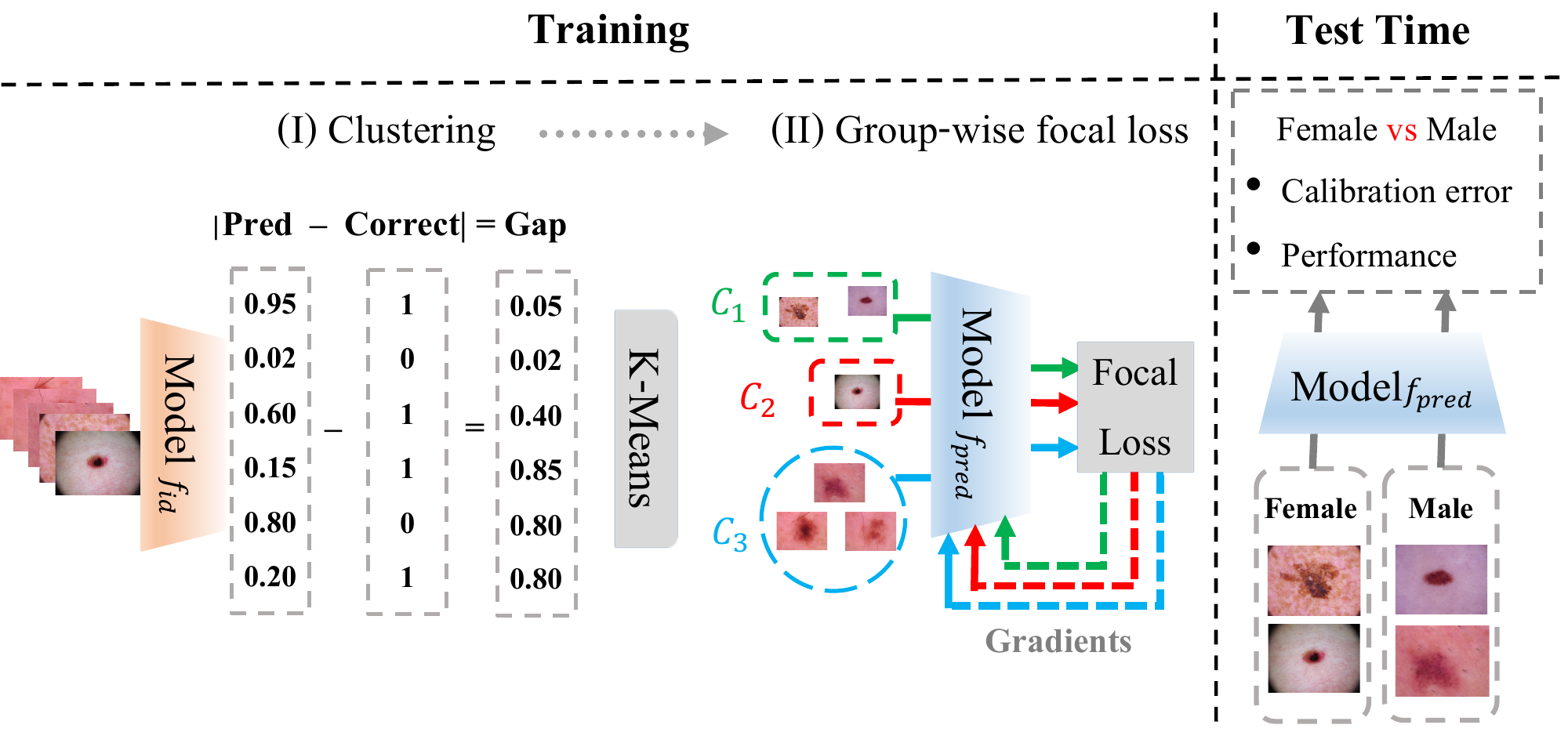}
  \caption{\textsf{Cluster-Focal} framework. The training procedure is a two-stage method, poorly calibrated sample identifications (clustering) and group-wise focal loss. At test time, the trained model $f_{\text{pred}}$ is deployed, then calibration bias and prediction performance are evaluated across various subgroup splittings such as sex or age. (Female/male patients are visualized as an example.) }
  \label{fig:protocol}
\end{figure}

We denote $D=\{(\x_i,y_i)\}_{i=1}^{N}$ as a dataset, where $\x_i$ represents multi-modal medical images and $y_i \in \{1, 2, \dots \}$ are the corresponding ground-truth class label. A neural network $f$ produces $\hat{p}_{i,y} = f(y|\x_i)$, the predicted probability for a class $y$ given $\x_i$. The predicted class for an $\x_i$ is defined as $\hat{y}_i = \text{argmax}_{y}~\hat{p}_{i,y}$, with the corresponding prediction confidence $\hat{p}_i = \hat{p}_{i,{\hat{y}_i}}$.

\subsection{Training procedure: two-stage method}
\paragraph{\textbf{Stage 1: Identifying poorly calibrated samples~(Clustering)}} In this stage, we first train a model $f_{\text{id}}$ via ERM~\cite{vapnik1991principles}, which implies training a model by minimizing the average training cross entropy loss, without any fairness considerations. $f_{\text{id}}$ is then used to identify samples that have potentially different calibration properties. Concretely, we compute the gap between prediction confidence $\hat{p}_i$ and correctness via $f_{\text{id}}$:
\begin{equation}
\label{eq:gap}
    \text{gap}(\x_i) = |\hat{p}_i - \mathbf{1}\{\hat{y}_i= y_i\}|,
\end{equation}
where $\hat{p}_i$ is the confidence score of the predicted class. Intuitively, if $\text{gap}(\x_i)$ is small, the model made a correct and confident prediction. When $\text{gap}(\x_i)$ is large, the model is poorly calibrated (i.e. incorrect but confident) for this sample. When the model makes a relatively under-confident prediction, $\text{gap}(\x_i)$ is generally in between the two values. We apply \emph{K-means} clustering  on the gap values, $\text{gap}(\x_i)$, to identify $K$ clusters ($C_1,\dots,C_K$) with different calibration properties. 

\paragraph{\textbf{Stage 2: Group-wise focal loss}} We then train a prediction model $f_\text{pred}$ with a group-wise focal loss on the clusters $C_1,\dots,C_K$ identified in the first stage. Formally, the following loss is used: 
\[
\calL_{\text{g-focal}} = \frac{1}{K}\sum_{k=1}^{K} \calL_{C_k}(f_\text{pred}),
\]
where $\calL_{C_k}(f_\text{pred}) = -\E_{(\x_i,y_i)\sim C_k}\left[\left(1-f_{\text{pred}}(y_i|\x_i)\right)^{\gamma}\log(f_{\text{pred}}(y_i|\x_i))\right]$ with $\gamma>0$. Intuitively, the focal loss penalizes confident predictions with an exponential term $\left(1-f_{\text{pred}}(y_i|\x_i)\right)^{\gamma}$, thereby reducing the chances of poor calibration~\cite{mukhoti2020calibrating}. Additionally, due to clustering based on $\text{gap}(\x_i)$, poorly calibrated samples will end up in the same cluster. The number of samples in this cluster will be small compared to other clusters for any model with good overall performance. As such, doing focal loss separately on each cluster instead of on all samples will implicitly increase the weight of poorly calibrated samples and help reduce bias.
 
\subsection{Test time evaluation on subgroups of interest}

\noindent At test time, we aim to mitigate the calibration error for the \textbf{worst-performing subgroup} for various subgroups of interest~\cite{diana2021minimax}. For example, if we consider sex (M/F) as the sensitive attribute and denote $\text{ECE}_{A=M}$ as the expected calibration error (ECE) on male patients, then the worst-performing subgroup ECE is denoted as $\max(\text{ECE}_{A=F},\text{ECE}_{A=M})$. Following the strategy proposed in \cite{nixon2019measuring,roelofs2022mitigating}, we use  Q(uantile)-ECE to estimate the calibration error, an improved estimator for ECE that partitions prediction confidence into discrete bins with an \emph{equal number of instances} and computes the average difference between each bin’s accuracy and confidence. 


In practice, calibration performance cannot be considered in isolation, as there  always exists a \emph{shortcut} model that can mitigate calibration bias but have poor prediction performance, e.g, consider a purely random (under-confident) prediction with low accuracy. As such, there is an inherent \textbf{trade-off} between  calibration bias and prediction error. When measuring the effectiveness of the proposed method, the objective is to ensure that calibration bias is mitigated without a substantial increase in the prediction error.

\section{Experiments and Results}\label{sec:exp}
Experiments are performed on two different medical image analysis tasks. We evaluate the performance of the proposed method against popular debiasing methods. We examine whether these methods can mitigate calibration bias without severely sacrificing performance on the worst-performing subgroups. \\

\begin{figure}[t]
\centering
\begin{subfigure}{.45\textwidth}
  \centering
  \includegraphics[width=.9\linewidth]{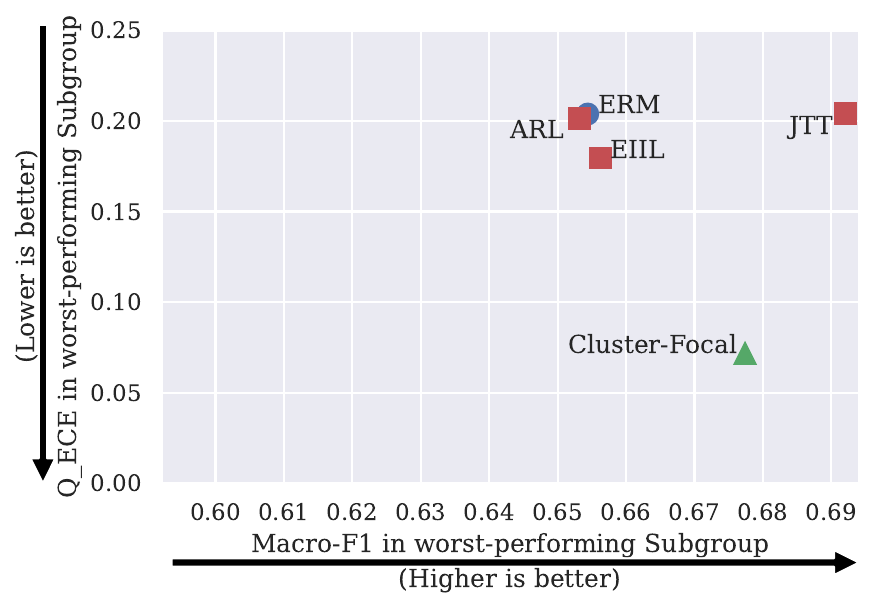}
  \caption{Age ($\leq 60$, $>60$)}
\end{subfigure}%
\begin{subfigure}{.45\textwidth}
  \centering
  \includegraphics[width=.9\linewidth]{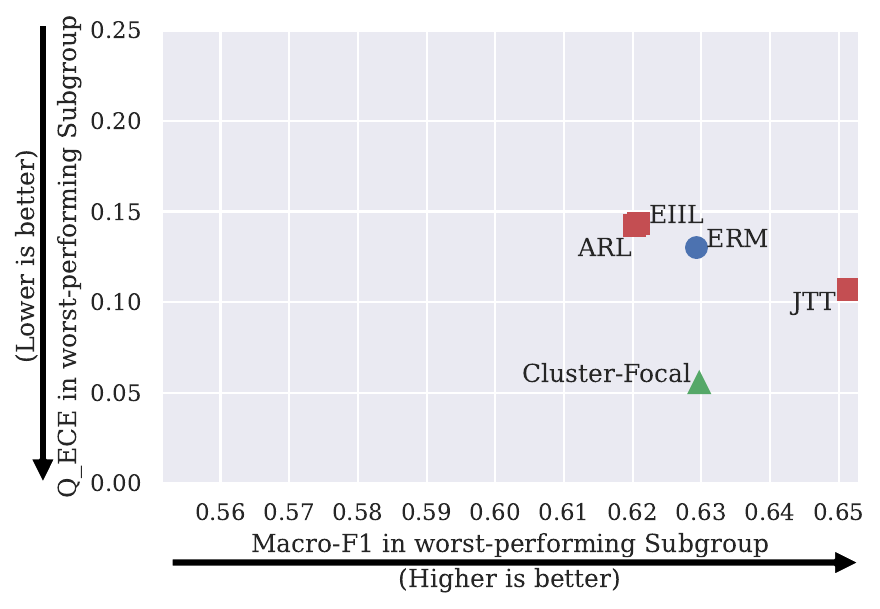}
  \caption{Sex (male, female)}
\end{subfigure}
\caption{HAM10000: worst performing subgroup results. \textsf{Cluster-Focal}: Proposed method; \textsf{ERM}: Vanilla model; \textsf{EIIL, ARL, JTT}: Bias mitigation methods. \textsf{Cluster-Focal} demonstrates a better trade-off, significantly improving worst-performing calibration with only a small degradation in prediction performance.}
\label{fig:trade_off_ham10k}
\end{figure}

\noindent\textbf{\underline{Task 1}: Skin lesion multi-class (n=7) classification.} HAM10000 is a public skin lesion classification dataset containing 10,000 photographic 2D images of skin lesions. We utilize a recent MedFair pipeline~\cite{zong2023} to pre-process the dataset into train (80\%), validation (10\%) and test (10\%) sets. Based on the dataset and evaluation protocol in \cite{zong2023}, we test two demographic subgroups of interest: age ($\text{age}\leq 60$, $\text{age}>60$), and sex (male, female).

\noindent\textbf{\underline{Task 2}: Future new multiple sclerosis (MS) lesional activity prediction (binary classification).} We leverage a large multi-centre, multi-scanner proprietary dataset comprised of MRI scans from 602 RRMS (Relapsing-Remitting MS) patients during clinical trials for new treatments~\cite{ms_advance_trial,ms_define_trial,ms_bravo_trial}. The task is to predict the (binary) presence of new or enlarging T2 lesions or Gadolinium-enhancing lesions two years from their current MRI. The dataset was divided as follows: training (70\%) and test (30\%) sets, validation is conducted through 4-fold cross validation in training set.  We test model performance on four different subgroups established in the MS literature
\cite{lampl2012weekly,lampl2013efficacy,signori2015subgroups,devonshire_relapse_2012,simon_ten-year_2015}. This includes: age ($\text{age} < 50$, $\text{age}\geq50$), sex (male, female), T2 lesion volume ($\text{vol} \leq 2.0 \text{ml}$, $\text{vol} > 2.0\text{ml}$) and Gad lesion count ($\text{count} = 0$, $\text{count} > 0$). Age and sex are sensitive demographic attributes that are common for subgroup analysis. The image-derived attributes were chosen because high T2 lesion volume, or the presence of Gad-enhancing lesions, in baseline MRI is generally predictive of the appearance of new and enlarging lesions in future images. However, given the heterogeneity of the population with MS, subgroups \emph{without} these predictive markers can still show future lesional activity. That being said, these patients can form a subgroup with poorer calibration performance. 

\begin{figure}[t]
\centering
\begin{subfigure}{.45\textwidth}
  \centering
  \includegraphics[width=.8\linewidth]{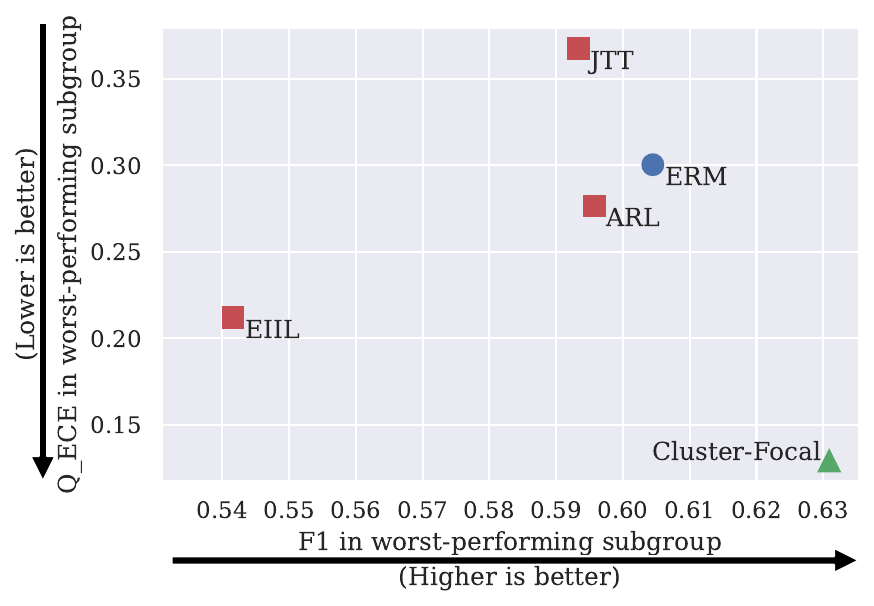}
  \caption{Age ($<50$, $\geq 50$)}
  \label{fig:sfig1}
\end{subfigure}%
\begin{subfigure}{.45\textwidth}
  \centering
  \includegraphics[width=.8\linewidth]{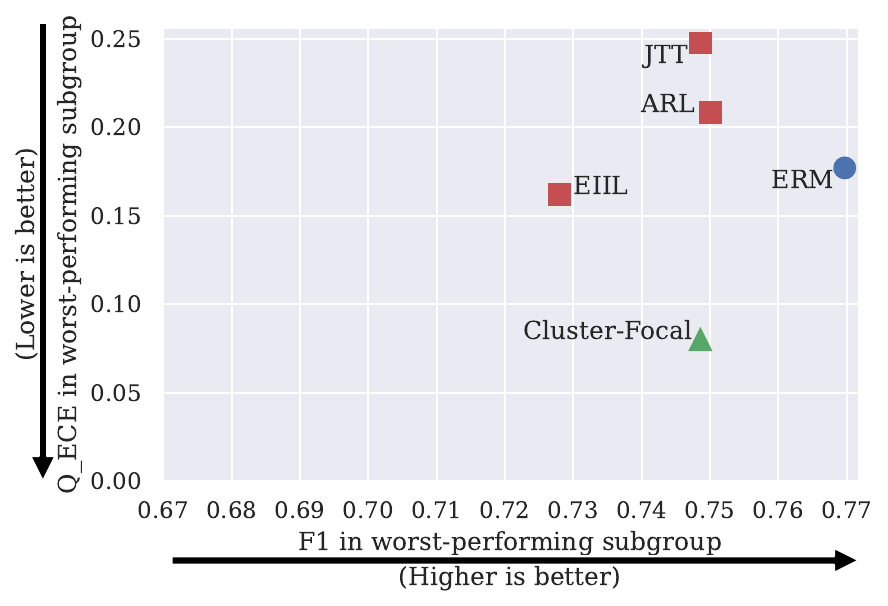}
  \caption{Sex (male, female)}
  \label{fig:sfig2}
\end{subfigure}
\begin{subfigure}{.45\textwidth}
  \centering
  \includegraphics[width=.8\linewidth]{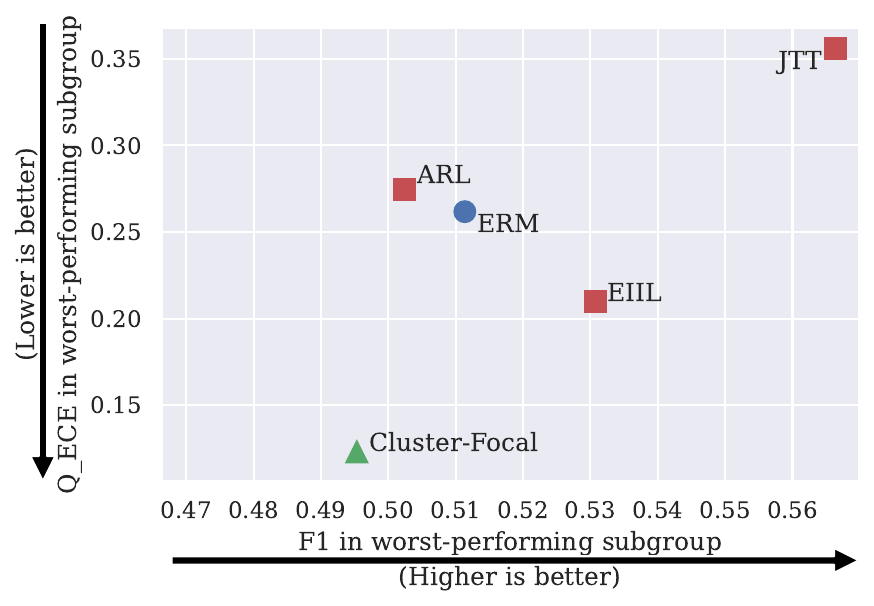}
  \caption{{T2 Lesion Volume ($\leq2\text{ml}$, $>2\text{ml}$)}}
\end{subfigure}
\begin{subfigure}{.45\textwidth}
  \centering
  \includegraphics[width=.8\linewidth]{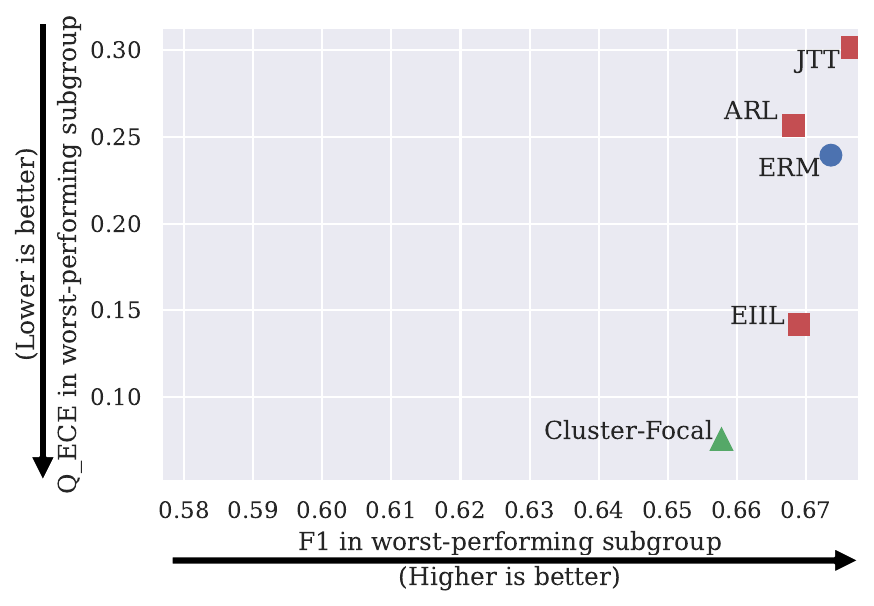}
  \caption{{GAD Lesion Count ($=0$, $>0$)}}
\end{subfigure}
\caption{MS: worst performing subgroup results. \textsf{Cluster-Focal}: proposed method; \textsf{ERM}: Vanilla Model; \textsf{EIIL, ARL, JTT}: bias mitigation methods.}
\label{fig:trade_off_MS}
\end{figure}

\noindent\textbf{\underline{Implementation Details}:} We adopt 2D/3D ResNet-18~\cite{he2016deep} for Task 1 and Task 2 respectively. All models are trained with Adam optimizer. Stage 1 model $f_{\text{id}}$ is trained for $10$ (Task 1) and $300$ (Task 2) epochs and Stage 2 prediction model $f_{\text{pred}}$ for $60$ (Task 1) and $600$ (Task 2) epochs. We set the number of clusters to 4 and $\gamma=3$ in group-wise focal loss. Averaged results across 5 runs are reported.

\noindent\textbf{\underline{Comparisons and Evaluations}:}~ Macro-F1 is used to measure the performance for Task 1 (7 class), and F1-score is used for Task 2 (binary). Q-ECE~\cite{mukhoti2020calibrating} is used to measure the calibration performance for both tasks. The performance of the proposed method is compared against several recent bias mitigation methods that do not require training with subgroup annotations: \textsf{ARL}~\cite{NEURIPS2020_07fc15c9}, which applies a min-max objective to reweigh poorly performing samples; \textsf{EIIL}~\cite{creager2021environment}, which proposes an adversarial approach to learn invariant representations, and \textsf{JTT}~\cite{liu2021just}, which up-weights challenging samples. Comparisons are also made against \textsf{ERM}, which 
trains model without any bias mitigation strategy. For all methods, we evaluate the trade-off between the prediction performance and the reduction in Q-ECE error for the \textbf{worst-performing subgroups} on both datasets.

\begin{figure}[t]
\centering
\begin{subfigure}{.5\textwidth}
  \centering
  \includegraphics[width=.8\linewidth]{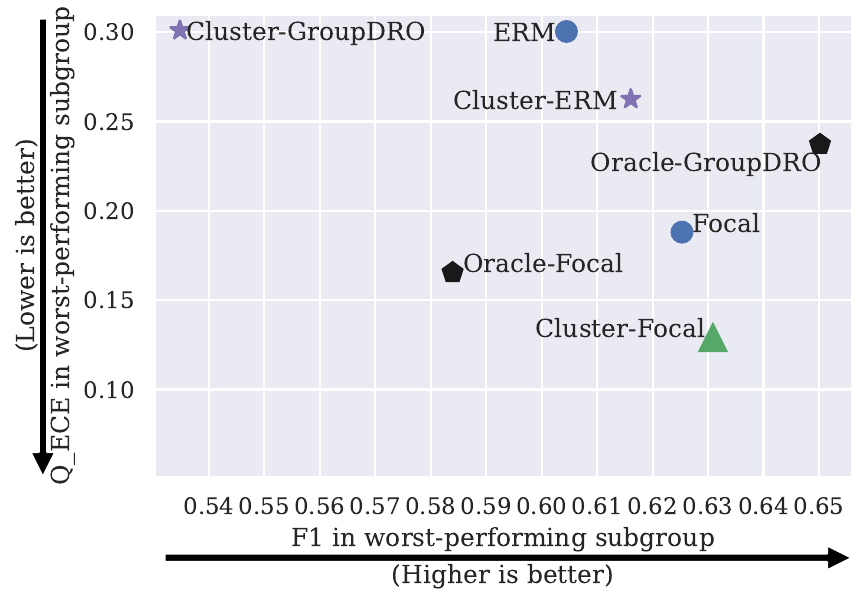}
  \caption{Age ($< 50, \geq 50$)}
\end{subfigure}%
\begin{subfigure}{.5\textwidth}
  \centering
  \includegraphics[width=.8\linewidth]{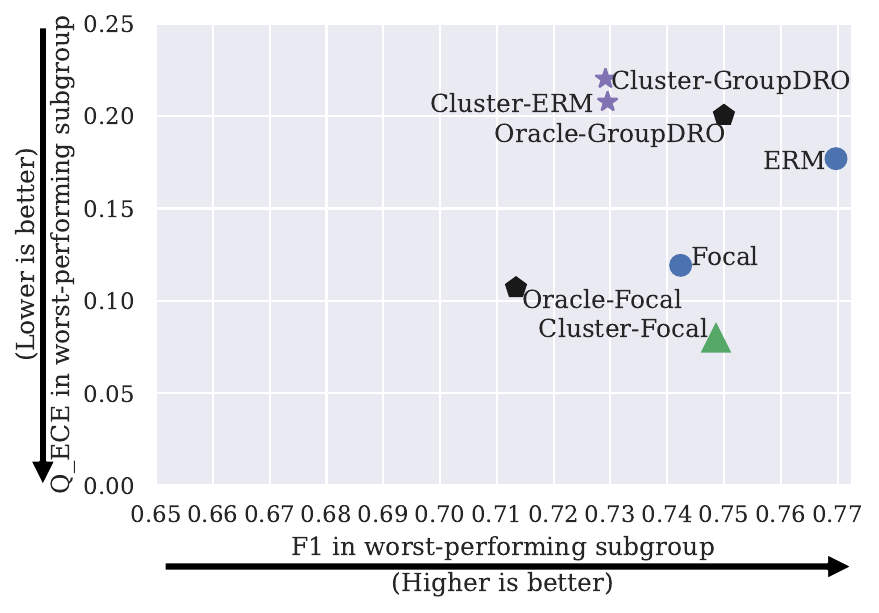}
  \caption{Sex (Male, Female)}
\end{subfigure}
\caption{Ablation Experiments for MS. \textsf{Focal}: regular focal loss without stage 1; \textsf{Cluster-ERM}: In stage 2, cross entropy loss is used; \textsf{Cluster-GroupDRO}: In stage 2, GroupDRO loss is used; \textsf{Oracle-Focal}: identified cluster in stage 1 is replaced by the subgroup of interest (oracle); \textsf{Oracle-GroupDRO}: GroupDRO method applied on the subgroups of interest.}
\label{fig:ablation_ms}
\end{figure}
\subsection{Results, ablations, and analysis} 
\noindent\textbf{\underline{Results}:} The resulting performance vs. Q-ECE errors tradeoff plots for worst-performing subgroups are shown in Fig.~\ref{fig:trade_off_ham10k} and \ref{fig:trade_off_MS}. The proposed method (\textsf{Cluster-Focal}) consistently outperforms the other methods on Q-ECE while having minimal loss in performance, if any. For instance, when testing on sex (male/female) for the MS dataset, (\textsf{Cluster-Focal}) loses around 2\% prediction performance relative to (\textsf{ERM}) but has around $8\%$ improvement in calibration error. When testing on sex in the HAM10000 dataset, we only observe a $2\%$ performance degradation with a $4\%$ improvement in Q-ECE.

In addition to subgroups based on sensitive demographic attributes, we investigate how the methods perform on subgroups defined on medical image-derived features.  In the context of MS, results based on subgroups, lesion load or Gad-enhancing lesion count are shown in Fig.~\ref{fig:trade_off_MS}(c-d). The proposed method performs best, with results that are consistent with demographic based subgroups. For Gad-enhancing lesion count, when compared with \textsf{JTT}, \textsf{Cluster-Focal} improves Q-ECE by $20\%+$ with a reduction in the prediction performance on the worst-performing subgroup of $2\%$. Detailed numeric values for the results can be found in the Supplemental Materials. 

\noindent\textbf{\underline{Ablation Experiments}:} Further experiments are performed to analyze the different components of our method. The following variant methods are considered: (1)~\textsf{Focal}: Removing stage 1 and using regular focal loss for the entire training set; (2)~\textsf{Cluster-ERM}: Group-wise focal loss in stage 2 is replaced by standard cross entropy; (3)~\textsf{Cluster-GroupDRO}: Group-wise focal loss in stage 2 is replaced by GroupDRO \cite{sagawadistributionally}; (4)~\textsf{Oracle-Focal}: In stage 1, the identified cluster is replaced by the true subgroups evaluated on at test time (oracle); (5)~\textsf{Oracle-GroupDRO}: We use GroupDRO with the true subgroups used at test time. Results for MS, shown in Fig.~\ref{fig:ablation_ms}, illustrate that each stage of our proposed model is required to ensure improved calibration while avoiding performance degradation for the worst-performing subgroups.

\begin{figure}[t]
\centering
\begin{subfigure}{0.45\textwidth}
    \centering
    \includegraphics[width=0.8\linewidth]{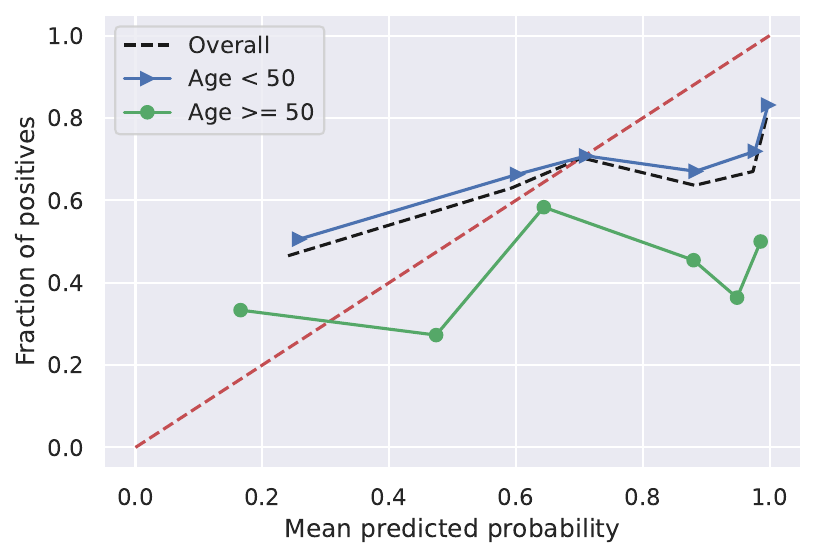}
    \caption{\textsf{EIIL}}    
\end{subfigure}
\begin{subfigure}{0.45\textwidth}  
    \centering 
    \includegraphics[width=0.8\linewidth]{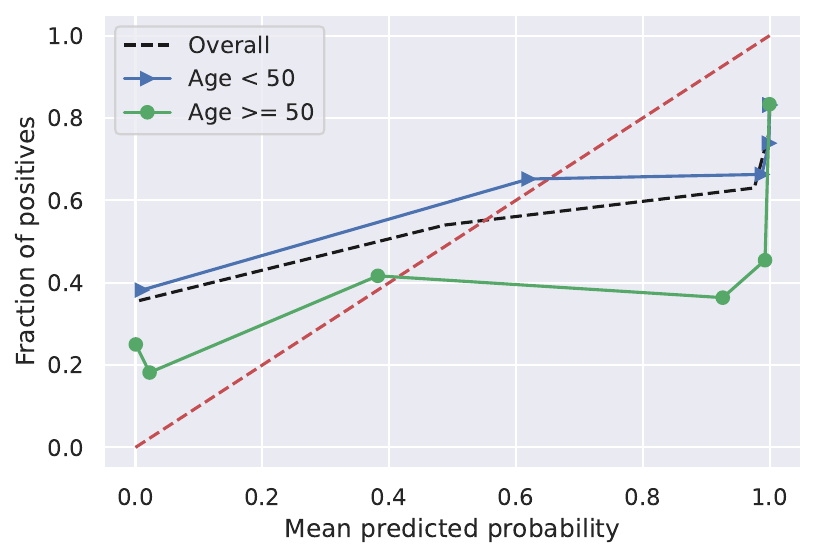}
    \caption{\textsf{ARL}}    
\end{subfigure}
\begin{subfigure}{0.45\textwidth}   
    \centering 
    \includegraphics[width=0.8\linewidth]{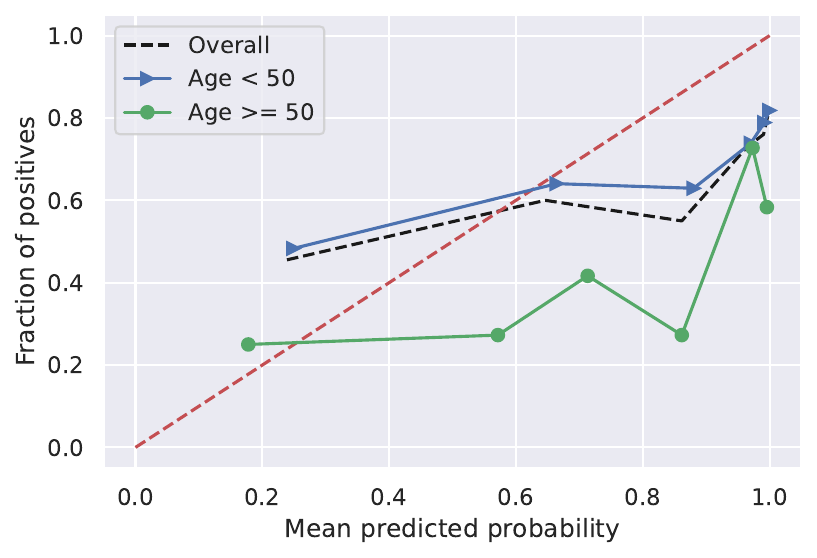}
    \caption{\textsf{JTT}}    
\end{subfigure}
\begin{subfigure}{0.45\textwidth}   
    \centering 
    \includegraphics[width=0.8\linewidth]{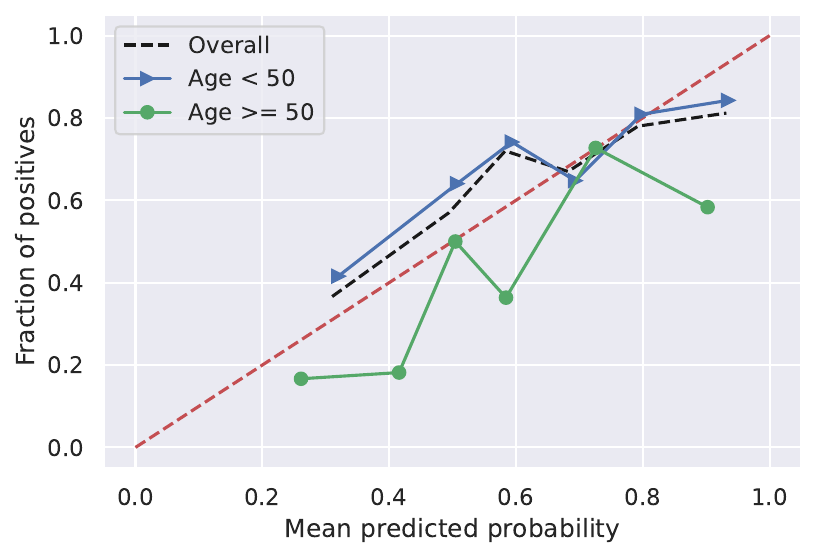}
    \caption{\textsf{Cluster-Focal}~(ours)}    
\end{subfigure}
\caption{MS: Reliability diagram for bias mitigation methods with age-based subgroups: (a) \textsf{EIIL}, (b) \textsf{ARL}, (c) \textsf{JTT}, and (d) \textsf{Cluster-Focal}.} 
\label{fig:calibration}
\end{figure}

\noindent{\textbf{\underline{Calibration Curves}:}} Fig.~\ref{fig:calibration} shows the reliability diagram for competing methods on Task 2: predicting future new MS lesional activity, with age being the chosen subgroup of interest (also see Fig.~\ref{fig:motivation}(a) for \textsf{ERM} results). Results indicate that popular fairness mitigation methods are not able to correct for the calibration bias in older patients (i.e. the worst-performing subgroup). With ARL, for example, most of the predictions were over-confident, resulting in a large calibration error. In contrast, our proposed method (\textsf{Cluster-Focal}) could effectively mitigate the calibration error in the worst-performing subgroup. 

\section{Conclusions}
In this paper, we present a novel two stage calibration bias mitigation framework (\textsf{Cluster-Focal}) for medical image analysis that (1) successfully controls the trade-off between calibration error and prediction performance, and (2) flexibly overcomes calibration bias at test time without requiring pre-labeled subgroups during training. 
We further compared our proposed approach against different debiasing methods and under different subgroup splittings such as demographic subgroups and image-derived attributes. Our proposed framework demonstrates smaller calibration error in the worst-performing subgroups without a severe degradation in prediction performance. \\ \\ 
\noindent\textbf{Acknowledgements} This paper was supported by the Canada Institute for Advanced Research (CIFAR) AI Chairs program and the Natural Sciences and Engineering Research Council of Canada (NSERC). The MS portion of this
paper was supported by the International Progressive Multiple Sclerosis Alliance (PA-1412-02420), the companies who generously provided the MS data: Biogen, BioMS, MedDay, Novartis, Roche/Genentech, and Teva, Multiple Sclerosis Society of Canada, Calcul Quebec, and the Digital Research Alliance of Canada.

%
%
\bibliographystyle{splncs04}
\bibliography{my}

\end{document}